# A Unified Fault Ride Through Technique for Virtual Oscillator based Grid Forming Controllers

Ritwik Ghosh, Minghui Lu, Member IEEE
Emails: ritwikghoshlives@gmail.com, minghui.lu@pnnl.gov

***Abstract*— Grid-forming technology has a crucial role in achieving the future all renewable power grid. Among different types of grid-forming controllers, Virtual Oscillator (VO) based Controllers (VOCs) are the most advanced. VOCs outperform the conventional droop-based grid-forming controllers in terms of dynamic performance and synchronization stability by adapting time-domain-based implementation. However, because of the time-domain-based implementation, the same Fault Ride-through (FRT) techniques for droop-based controllers are incompatible with VOCs. Existing literature has successfully incorporated current limiting techniques in VOCs to protect the converters during severe transient conditions. Nevertheless, some very important aspects of FRT requirements are not attended to by the existing literature on VOCs, such as maintaining synchronization with the network during a fault, minimizing power oscillation during a fault, and at the fault clearance. First, this article introduces a unique analytical approach for quantifying the underlying dynamics of VOCs during faults. Next, using the mentioned analysis and in-depth reasoning, the systematic development of a unique FRT control architecture for VOCs is presented. The proposed FRT technique has unified both current and voltage synchronization in the same architecture to work successfully under three-phase and unbalanced faults. The performance of the proposed controller is thoroughly investigated and compared with existing VOCs.**

## I. Introduction

Inverter-based Resource (IBR) is one of the most vital links between the electrical power grid and the Renewable Energy Sources (RESs) [1]. In the near future, it is expected that a significantly large portion of the energy of the electrical power grid will be fed through IRRs [1]. Newly introduced standards for the interconnection of IBRS with the power grid, e.g., [2] are tightening the requirements for important functionalities such as frequency support, voltage support, and fault ride-through capabilities in IBRs. Grid-forming control strategies are enabling these crucial features in IBRs.

Among three types of grid-forming controllers, i.e., the droop controller, the virtual synchronous machine controller, and the Virtual Oscillator based Controller (VOC), the VOC is the latest and most advanced [3]. VOCs use time-domain mathematical model of non-linear oscillators to control grid-forming inverters [3], [4]. By adopting the mentioned time-domain implementation, VOCs provide better dynamic response than conventional droop and virtual synchronous machine controllers [5], [6]. At the same time, VOCs provide all the steady-state functionalities of conventional droop and virtual synchronous machine controllers [6]. In addition, the latest VOC, i.e., Andronov-Hopf oscillator based controllers, can achieve harmonic-free output voltages without compromising the dynamic response [3], [7].

In the last decade, researchers have accomplished remarkable success in the development and advancement of VOCs. Different non-linear oscillator models, such as dead zone [8], Van-der-pol [4], modified Van-der-pol [9], and Andronov-Hopf [7] are proposed by the researchers to construct grid-forming controllers. Finally, it has been proved that the Andronov-Hopf oscillator model is the best fit for grid-forming controllers [3]. Crucial functionalities such as grid compatibility [7], [10], MPPT [11], working under unbalanced conditions [12], and current control [13] are enabled in VOCs. System-level control architectures are developed for VOCs. Despite the mentioned notable developments, one crucial functionality, i.e., fault ride-through capability, demands significant improvements to enable widespread usability of VOCs. A simulation study in Section II shows that the most advanced system-level VOC fails to meet crucial fault ride-through functionalities such as maintaining synchronization with the network, preventing active power swing and reversal, and staying connected to the network.

Any grid-forming source is controlled to mimic the steady-state and dynamic characteristic of Synchronous Generators (SGs), even having a much lower thermal capacity than SGs [14], [15]. However, the recent grid codes enforce the grid-forming sources to stay connected to the grid under fault conditions when the transient current may tend to reach up to four or five times the rated value [15], [16]. As a result, any grid-forming controller, including VOC, must activate current control mode in order to limit the output current during fault conditions [17], [18], [19]. Interestingly, current limiting is just one part of the Fault Ride Through (FRT) operation to keep a GFM connected to a network during a fault [20]. There are other crucial parts of FRT operations, such as maintaining effective synchronization, preventing active power swing and reversal, maintaining normal operation at the healthy phases, and reactive power support.

A few control approaches are presented in existing literature for enabling fault ride-through operation in VOCs. A fault ride-through control technique is proposed for a component-level dispatchable VOC (d-VOC) in [13], [21]. In the mentioned method, the main component-level model of a dispatchable virtual oscillator is modified by adding a Virtual Impedance (VI) loop to limit the output current under fault conditions. The mentioned approach has critical drawbacks. Firstly, because it modifies the virtual oscillator model itself, the mentioned technique is not directly compatible with any latest system-level control architecture for VOCs, such as presented in [11], [22], [23], where inner nested voltage and current control loops are added to a component-level virtual oscillator model. Next, in the mentioned modified d-VOC model, the zero-sequence voltage component is ignored, which limits its operation in three-phase four-wire systems. It is important to mention that, as presented in [12], zero sequence voltage synchronization plays a crucial role for a VOC to perform under unbalanced conditions in three-phase four-wire systems. Since the FRT controller and the d-VOC model are not decoupled in [13], [21], newly proposed advanced concepts for FRT, such as adaptive synchronization technique during fault conditions [24], cannot be added to the existing control architecture from the top. Current limiters and anti-windup loops are incorporated into the inner nested voltage controllers to limit output current under fault conditions in the system-level VOC presented in [11], [22]. However, as it is observed from the simulation study presented in Section II, due to the interaction between the current limiters and the virtual oscillator, the VOC loses effective synchronization with the network during the fault condition. As a result, the grid-forming source experienced severe active power swings and active power reversal, which led to an unsuccessful FRT operation. An estimated current synchronization technique is presented in [23] for system-level VOC to enable FRT operation. However, the FRT controller presented in [23] needs at least one healthy phase to ensure effective synchronization with the connected network under fault conditions. It is essential to mention that the non-linear reactive power droop characteristics prevent VOC from maintaining effective synchronization with a connected network if the voltage amplitude of the PCC drops more than 15% under any transient condition [25]. At the same time, due to its time domain implementation, which is different from the phasor-based implementation of conventional droop controllers, the existing FRT techniques for conventional droop controllers are not compatible with VOCs.

This article focused on addressing the limitations of FRT controllers for VOCs mentioned above. The unique research contributions of this article are summarized as follows.

**1.** First, this article has presented a simulation study that identifies the exact problem faced by existing VOCs under fault conditions. It has been seen that the existing VOCs can successfully limit the output current during faults. However, effective synchronization is lost due to the interaction between the current limiters and the virtual oscillator. The loss of synchronization results in active power swing and reversal and, ultimately, disconnection of the grid-forming source from the network.

**2.** Next, a unique analytical method is presented for mathematically quantifying the underlying dynamics of a VOC under fault conditions. The analysis mathematically expresses the interaction between the current limiters and the virtual oscillator during a fault condition. Extending the mentioned mathematical expressions, specific reasons are identified that cause a VOC to go out of synchronization, and to experience active power swing and reversal during a fault condition.

The proposed mathematical method can also be used to analytically predict the states of important performance parameters such as load angle and active power under the corner fault cases. It can be used to analytically validate the performance of any new FRT controller for VOCs.

**3.** Using the above-mentioned mathematical analysis, this article has derived the general criteria for the successful FRT operation of any VOCs.

**4.** Finally, a unified fault ride-through controller for VOCs is presented. The systematic development of the proposed controller is presented in this article. The proposed controller has successfully unified current and voltage synchronization in the same system-level architecture to work under balanced and unbalanced faults. The proposed FRT controller can be incorporated into the existing VOCs without the need for any changes in the existing control architecture or extra sensors.

**5.** The performance of the proposed FRT controller is thoroughly investigated. A performance comparison is presented between an existing VOC and the same VOC integrated with the proposed FRT controller. The results show that the proposed FRT controller improves the performance of the existing VOC significantly and helps to meet the critical FRT requirements, such as maintaining effective synchronization, preventing active power swing and reversal, maintaining normal operation at the healthy phases, and reactive power support.

The rest of the article is organized as follows. Section II has presented a simulation study to analyze the performance of an existing VOC under fault conditions. A mathematical analysis is presented in Section III that quantify the underlying dynamics of a VOC under fault conditions and derive the criteria for successful FRT operation. The systematic development of the proposed Unified FRT controller for VOC is presented in Section IV. The performance of the proposed controller is thoroughly investigated in Section V. The article is finally concluded in Section VI.

## II. Loss of Effective Synchronization from Existing VOCs under a Fault

The loss of effective synchronization creates a severe problem for an existing VOC-controlled grid-forming source to stay connected to the network under fault conditions. A simulation study has been conducted to analyze the underlying dynamics that cause this problem. The schematic diagram and the system specifications used for the simulation study are presented in Fig. 1. and Table I, respectively.

It is essential to mention that the simulation study is conducted with a bidirectional voltage source. The intention is to observe and analyze the actual characteristics of an existing VOC. Next, the problems that will occur in an actual scenario when the voltage source is unidirectional, dispatching energy only from the source to the grid, are described in detail.

The three-phase power reference, $P^*$ is set to 9 kW. The upper and lower limits for the direct axis current, $I_{d_upper}$ and $I_{d_lower}$ are set to 0 and 20 A, respectively. At the time, $t_1 = 0.5$s a fault is created from the grid-side at the PCC. The fault is cleared at $t_1 = 0.75$s. The voltage at the PCC, $v_{PCC}$, the output current of the grid-forming voltage source, $i_{inv}$, the three-phase power output of the grid-forming source, $P$ and the sine of the load angle between the grid-forming source and the grid, $sin\,\delta$ are presented in Fig. 2. The simulation study provides the following crucial observations that are used for analysis.

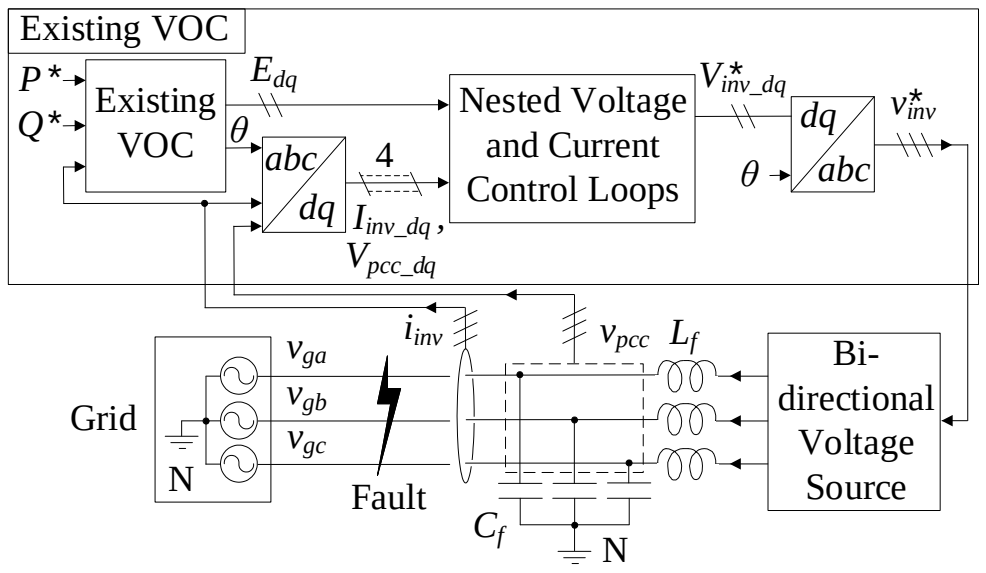

Fig. 1. Existing VO-based grid-forming system-level controller in the presence of an unbalanced fault

Table I: Specifications of the system used for the simulation

| Symbol | Description | Value |
|---|---|---|
| $V_n$ | Nominal grid voltage: Phase (rms) | 400 V |
| $\omega_n$ | Nominal frequency | $2\pi 50$ rad/s |
| $L_f$ | Filter inductor | 2 mH/Phase |
| $C_f$ | Filter capacitor | 8 μF/Phase |
| $I_{max}$ | Overcurrent limit: Phase (rms) | 20 A |

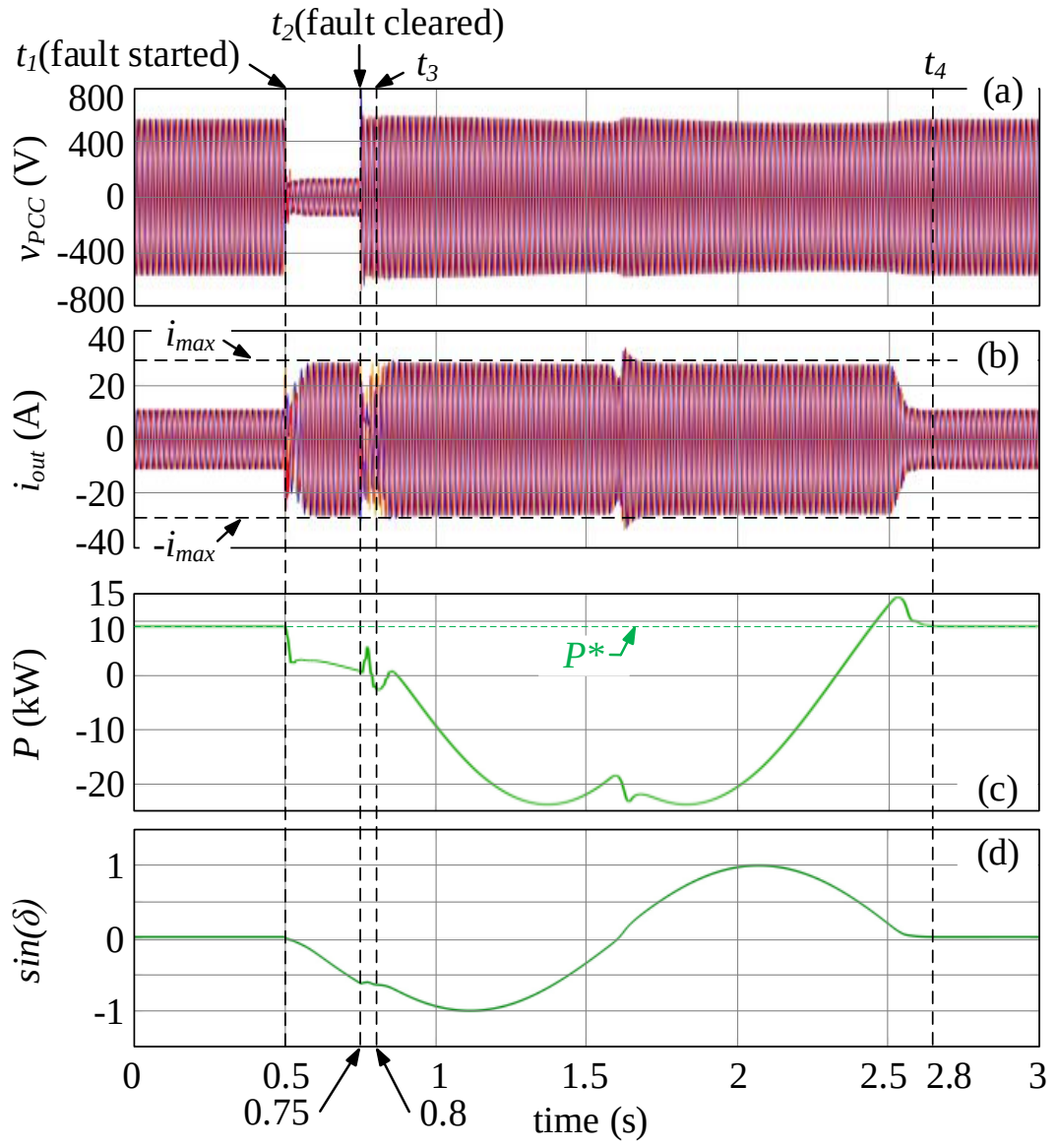

Fig. 2. The performance of an existing VO-based grid-forming system-level controller in the presence of an unbalanced fault

- Fig. 2. (b) shows that the nested voltage controller can keep the output current under the limit during the fault.
- The large deviation in $sin\,\delta$ (where $\delta$ is the load angle), as shown in Fig. 2. (d) indicates that the controller has lost effective synchronization with the network due to the fault. The lack of effective synchronization causes the large active power swing, which can be observed from Fig. 2. (c). Interestingly, the active power output went negative, i.e., the grid-forming source started consuming power from the grid even though the limits of direct axis current, $I_{d_upper}$ and $I_{d_lower}$ were set from 0 and 20 A due to the loss of synchronization.
- The simulation results validate a crucial and advantageous feature of a Virtual Oscillator based controller. The feature is that a non-linear virtual oscillator, one that is used in the controller, can achieve effective and stable synchronization, starting from any arbitrary condition. Because of the mentioned feature, even after the load angle and power swing, the controller successfully restored a stable load angle and active power output at the time, $t_4$ as shown in Fig. 2.
- However, even with the advantageous features mentioned above, the controller will fail in an actual situation for the following reasons. A bi-directional grid-forming source is used in this simulation to analyze the complete dynamic response of the controller during a fault. Nevertheless, in actual situations, the grid-forming source will be unidirectional to dispatch energy from the source to the grid. Hence, in practical situations at the time, $t_3$ (as depicted in Fig. 2) the inverter will be disconnected from the grid when active power tends to reverse.

## III. Analysis

Prior to the occurrence of the fault, the grid-forming source was dispatching the active power, $P$ to the grid, which is very close to the reference power, $P^*$. The active and reactive power droop characteristic of the source is governed by the VOC and expressed as follows.

$$\omega_{VOC} = \omega_n - \frac{k_v k_i}{3CV_{VOC}{}^2}(P - P^*) \quad (1)$$

$$0 = \frac{\xi}{k_v^2} V_{VOC}(2V_n^2 - 2V_{VOC}^2) - \frac{k_v k_i}{3CV_{VOC}}(Q - Q^*) \quad (2)$$

The active power droop characteristic is the key to analyzing the reason for the loss of effective synchronization. The analysis should be started by presenting a brief overview of the frequency support functionality of a VOC considering an under-frequency situation.

The output power of the converter, $P$ follows the droop characteristic presented by (*1*). When the system frequency, $\omega_g$ and frequency of the VOC, $\omega_{VOC}$ are both equal to the nominal system frequency, $\omega_n$ the output power of the inverter, $P$ converges to the reference power, $P^*$. Now, at this point, if the system frequency is decreased, the internal load angle, $\delta$ of the inverter starts increasing because $\omega_{VOC} > \omega_g$. As a result, the inverter starts dispatching more power than the reference, $P^*$ which starts bringing down the frequency of the VOC, $\omega_{VOC}$ obeying the droop, (*1*). At the same time, when all the grid-forming sources start injecting more power into the system, the system frequency, $\omega_g$ start increasing. Finally, the system's frequency and the VOC converge to a new equilibrium. During an over-frequency event, the VOC reduces the power output and, in a similar fashion, converges to a new equilibrium with the system. Using this concept of droop, the underlying dynamics of a VOC during fault are analyzed below. The analysis follows the sequences of events that occurred during a fault.

***A.* <u>Activation of the Current Limiting Mode</u>:** Fig. 3 shows the phasor diagram that illustrates various voltage and current vectors at the start of the fault, $t = t_1$.

The voltage amplitude of the PCC and the VOC before the fault are denoted as $V_{PCCP}$ and $V_{VOCP}$ respectively. A positive load angle, $\delta_P$ was maintained by the VOC to dispatch active power to the connected network.

At the start of the fault, $t = t_1$, the voltage amplitude of the PCC is reduced from $V_{PCCP}$ to $V_{PCCF}$. The voltage amplitude of the controller is also reduced from $V_{VOCN}$ to $V_{VOCF}$, but as illustrated in Fig. 3, not at the same ratio as the voltage of the PCC. It is because of the non-linear reactive power droop characteristic of the VOC, as expressed in (2). Due to the mentioned non-linear reactive power droop characteristic, the voltage amplitude of a VOC, in any under-voltage situation, does not drop more than an amount of around 15% [25]. As a result, the current output of the grid-forming source, $I_{outP}$, tends to shoot up, which activates the current limiters and the anti-windup controller in the inner PI voltage control loop. Interestingly, the inner current loop is significantly faster than the active power time constant of the virtual oscillator, $\tau$ [11]. This is why, at the start of the fault, $t = t_1$ the output current reaches the saturation significantly faster than there occurs any change in the initial load angle, $\delta_P$. It is essential to mention that the grid-forming controller generates current references, $I_{dF}^*$ and $I_{qF}^*$ by taking the internal voltage of the virtual oscillator, $V_{VOCF}$ as the reference frame. The direct axis component of the output current, $I_{outF}$ with respect to the voltage of PCC, $V_{PCCF}$ is denoted by $I_{dPCCF}$ in Fig. 3. The direction of $I_{dPCCF}$ is positive at $t = t_1$. As a result, the active power output of the grid-forming source, $P$ is still positive at the start of the current limiting mode. However, the active power output, $P$ is reduced due to the reduction of the voltage amplitude of the PCC from $V_{PCCP}$ to $V_{PCCF}$. The instantaneous active power waveform in Fig. 2. (c) confirms the above-mentioned analysis about the output active power of the grid-forming source, $P$ at the start of the fault.

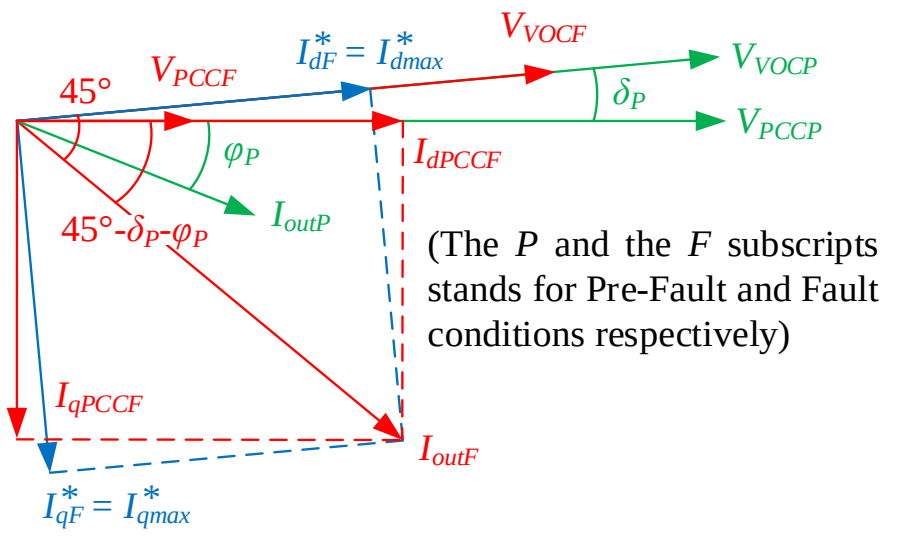

Fig. 3. The phasor diagram that illustrates various voltage and current vectors at the start of the fault

***B. The Decrease in Load Angle, $\delta_F$, During the Fault*:** A VOC achieves synchronization with a connected network by using only the instantaneous output current of the grid-forming source as feedback, as shown in Fig. 1. The voltage of the PCC is not used as feedback for the purpose of synchronization in a VOC. The power in (1) is estimated using the voltage amplitude of the VOC, $V_{VOC}$. The estimation works well under normal conditions when the difference between the voltage amplitudes of the VOC and the PCC is small, and the angle between the two voltage vectors is also small. However, at the fault condition, as mentioned in the last subsection, the voltage of the VOC differs significantly from the voltage of the PCC. As a result, (1) cannot provide a correct estimation of angular frequency and, hence, the load angle information during a fault. The angular frequency of the VOC during the fault, $\omega_{VOCF}$ is derived from (1) as follows by replacing the power terms with currents.

$$\omega_{VOCF} = \omega_P - \frac{k_v k_i}{2CV_{VOCF}}\left(I_{doutF} - I_{dVOCF}^*\right) \quad (3)$$

The angular frequency of the connected system prior to the fault was $\omega_P$. $I_{doutF}$ is the direct axis current output of the grid-forming source during the fault. The internal direct axis current reference of the VOC during the fault is $I_{dVOCF}^*$.

Now, during the fault, the output current of the grid-forming source is increased and saturated to the maximum current limit.

$$I_{doutF} = I_{dmax} \quad (4)$$

The internal current reference of the VOC during the fault, $I_{dVOCF}^*$ represented as

$$I_{dVOCF}^* = \frac{2P^*}{3V_{VOCF}} \quad (5)$$

is also increased. However, as mentioned in the last subsection, the voltage amplitude of the VOC during a fault, $V_{VOCF}$ does not decrease significantly from its nominal value, $V_n$ [25]. As a result, term $I_{dVOCF}^*$ does not increase significantly and remains smaller than $I_{doutF}$ during the fault.

$$I_{doutF} > I_{dVOCF}^* \quad (6)$$

Now, from (3) and (4), it is clear that the angular frequency of the VOC starts reducing during the fault. As a result, the load angle delta, which is represented as

$$\delta_F = \int (\omega_{VOCF} - \omega_P)\, dt + \delta_P$$
$$= \int -\frac{k_v k_i}{2CV_{VOCF}}\left(I_{dmax} - I_{dVOCF}^*\right) dt + \delta_P \quad (7)$$

starts decreasing and going towards the negative.

***C. The Active Power Reversal*:** As shown in Fig. 2. (c) the active power is reversed after the fault clearance. The mathematical analysis in this subsection shows that the interaction between the virtual oscillator and the inner PI voltage control loop plays a crucial role in the reversal of the active power to the grid-forming source and, hence, in the FRT performance. The analysis provides the limiting conditions for a VOC to achieve a successful fault ride-through operation.

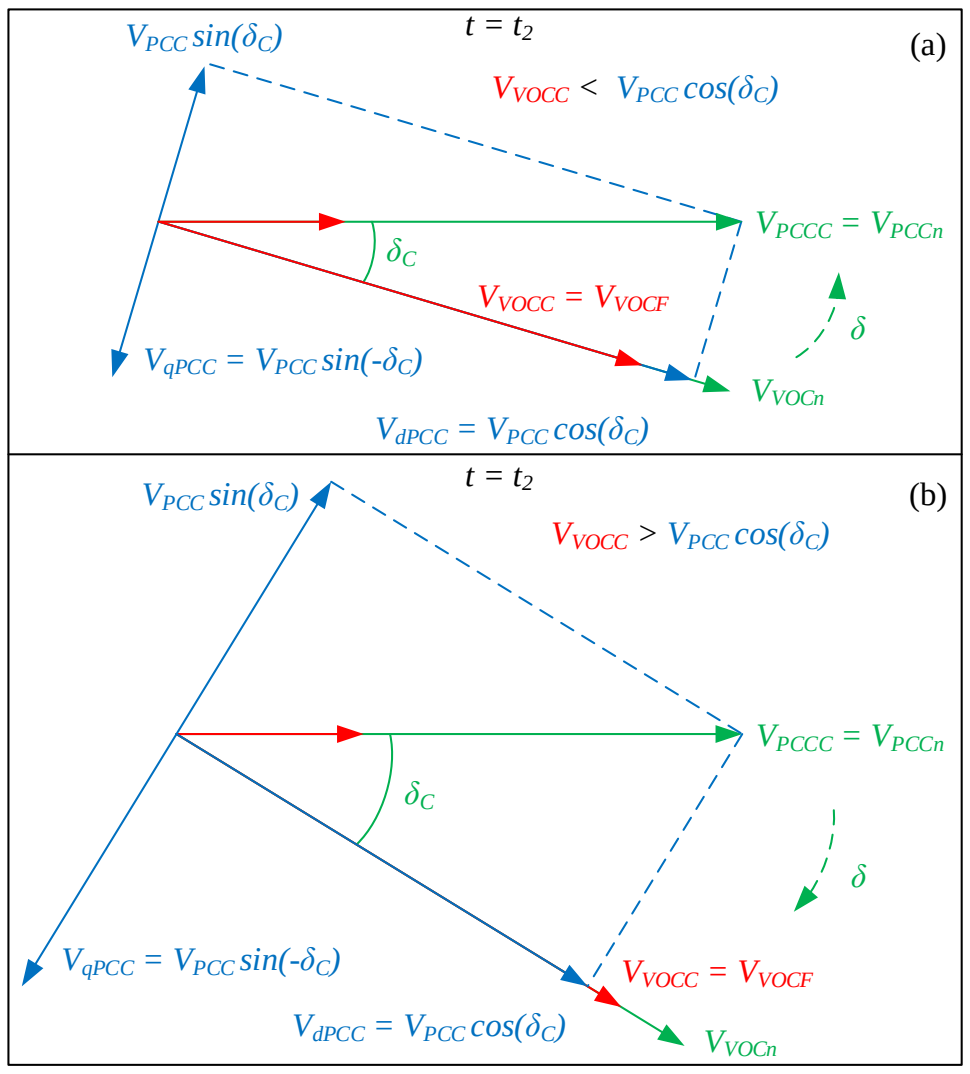

Fig. 4. The phasor diagram that illustrates various voltages and currents at the start of the fault

Fig. 2. (d) shows that, at the fault clearance, $t = t_2$ the VOC ends up with a load angle, $\delta_C$, which is negative with respect to the voltage of the PCC, $V_{PCC}$. The voltage amplitude of the connected network, $V_{PCC}$ immediately returns to the nominal value, $V_{PCCn}$ from $V_{PCCF}$. However, the voltage of VOC, $V_{VOC}$ takes time, depending on the voltage rise time of the virtual oscillator, $t_{rise}$ to reach back to $V_{VOCn}$ from $V_{VOCF}$. The value of $\delta_C$, can be calculated using (7). Now, at this instant, i.e., $t = t_2$ the fault is cleared, and as shown in Fig. 2. (b) the output current amplitude of the grid-forming source, $I_{out}$ starts reducing from the maximum rated current, $I_{max}$ for a short

duration. It indicates that the current saturation limiters and the anti-windup controller in the inner PI voltage control loop are deactivated.

As soon as the current saturation limiters are deactivated, the inner PI current loop starts tracking the direct and quadratic axis reference currents, $I_{dPI}^*$ and $I_{qPI}^*$ generated by the inner PI voltage controllers. The detailed equations of the inner PI voltage and current controllers are presented in [23] Using the values of voltage vectors illustrated in Fig. 4, the value of $I_{dPI}^*$ can be presented as follows.

$$I_{dPI}^* = \left(K_{pv} + \int K_{iv}\right)(V_{VOC} - V_{PCC}\cos\delta_C) \\ -\omega_n C_f V_{PCC}\sin(-\delta_C) \quad (8)$$

The terms, $K_{pv}$ and $K_{iv}$ represent the proportional and integral gain of the inner PI voltage controller, respectively. The nominal angular frequency of the system and filter capacitor are denoted by $\omega_n$ and $C_f$ respectively.

Next, considering the inner PI current controller efficiently tracks the direct axis current reference, the load angle after the fault clearance can be presented as follows by using the term, $I_{PId}^*$ in (7).

$$\delta = \int -\frac{k_v k_i}{2CV_{VOCF}}(I_{dPI}^* - I_{dVOC}^*)\,dt + \delta_C \quad (9)$$

The direct axis component of the current references generated by the virtual oscillator and inner PI voltage controller are denoted by $I_{dPI}^*$ and $I_{dVOC}^*$ respectively.

The behavior of a VOC after the fault clearance can be analyzed using (8) and (9). The two possible conditions after the fault clearance are presented below.

**(i) Condition 1: $I_{dPI}^* < I_{dVOC}^*$ at the fault clearance:** As shown in Fig. 4. (a) at the instant of fault clearance, $t = t_2$, if the load angle, $\delta_C$ is smaller in the negative direction, the voltage component of the PCC in the direction of $V_{VOC}$, $V_{PCC}\cos(\delta_C)$ become larger than $V_{VOCC}$, i.e., $V_{PCC}\cos(\delta_C) > V_{VOCC}$. As a result, the direct axis current reference from the inner PI voltage controller, $I_{dPI}^*$ which is represented mathematically in (8), tends to become smaller than the direct axis current reference generated by the virtual oscillator, $I_{dVOC}^*$, i.e., $I_{dPI}^* < I_{dVOC}^*$. Now, at this instant, if, $I_{dPI}^* < I_{dVOC}^*$, it can be seen from (9) that the magnitude of the load angle, $\delta$ starts reducing and moving towards a positive value from negative. The mentioned change in load angle, $\delta$ reduces the value of the terms, $(V_{VOC} - V_{PCC}\cos\delta_C)$ and $-\omega_n C_f V_{PCC}\sin(-\delta_C)$ in (8), which works against the further increase of $I_{PId}^*$. Again, because of a lower value of $I_{PId}^*$ than $I_{dVOC}^*$ helps the load angle, $\delta$ to move further towards positive value from negative. This self-reinforcing process continues until the load angle, $\delta$ converges to the pre-fault load angle value, $\delta_P$.

**(i) Condition 2: $I_{dPI}^* > I_{dVOC}^*$ at the fault clearance:** The Condition 2 occurred during the simulation study presented in Section II. At the end of the fault, $t = t_2$, the VOC ends up with a large negative load angle, $\delta_C$. It leads just the opposite to Condition 1, i.e., $V_{PCC}\cos(\delta_C) < V_{VOCC}$ and hence, $I_{dPI}^* > I_{dVOC}^*$. All the voltage and current vectors for this instant are illustrated in Fig. 4. (b).

The condition as mentioned above initiated a self-reinforcing act for the load angle, $\delta$ which happened in this simulation as follows.

**(a)** The load angle, $\delta$ tends to go further negative as per (8), as $I_{dPI}^* > I_{dVOC}^*$ which is a result of $V_{PCC}\cos(\delta_C) < V_{VOCC}$.

**(b)** As shown in Fig. 4. (b) the increment of $\delta$ towards a negative direction reduce the value of $V_{PCC}\cos(\delta_C)$ further and at the same time, the value of $V_{VOCC}$ increases towards $V_{VOCn}$. According to (8) it results in a further increase of $I_{dPI}^*$ than $I_{dVOC}^*$, which finally, as per (9) pushes the $\delta$ towards a further negative direction.

The two self-reinforcing acts written in (a) and (b) increase the value of $I_{dPI}^*$ until it hits back again to $I_{max}$. As a result, as shown in Fig. 2. (b) the output currents saturate back to the maximum current after a short duration of the fault clearance, $t = t_2$. The value of $\delta$ also continued to increase in the negative direction until it took a complete $360^\circ$ flip in the clockwise direction before finally converging to the pre-fault value $\delta_P$. In this process, the active power output becomes negative, as shown in Fig. 2. (c) before finally reaching back to the active power reference, $P^*$.

## IV. The Unified FRT Control Architecture for VOCs

The mathematical analysis presented in Section III provide the general criteria for virtual oscillator based grid-forming controllers to achieve successful fault ride-through operation. The mentioned criteria are given as follows.

**1.** During the fault, the load angle, $\delta_F$ needs to be kept as close as possible to the pre-fault load angle, $\delta_P$.

The simulation study of Section II and the analysis of Section III shows that the VOC cannot meet the mentioned criteria using the normal current feedback for synchronization during fault.

**2.** At the fault clearance, if the load angle, $\delta$ is negative the amplitude of the current reference applied to the inner PI current control loop, $I_{PId}^*$ needs to be smaller than the amplitude of the current reference applied to the virtual oscillator, $I_{dVOC}^*$, i.e., $I_{PId}^* < I_{dVOC}^*$.

The mentioned criteria must be maintained from the instant when the fault is cleared until the voltage amplitude of the VOC, $V_{VOC}$ recovers from the lower fault value, $V_{VOCF}$ to the pre-fault nominal value, $V_{VOCn}$.

Any Fault Ride-Through (FRT) controller for VOCs must satisfy the above two criteria. This section has presented the systematic development of a Unified FRT controller for VOCs.

***A.* Integration of the FRT Control Loop into the Existing System-level VOC:** The proposed FRT controller can be integrated into any existing system-level VOC [11], [22], [23] as it does not impose any dependency on the existing grid-forming control architecture and does not demand any extra sensors. Fig. 5 illustrates the complete virtual oscillator-based grid-forming control architecture with the integrated proposed unified FRT controller. The existing VOC [23] and the proposed FRT controller are depicted in black and red, respectively. A Symmetrical Component based Virtual Oscillator (S-VO) [12] is used to enable unbalanced operation. The proposed FRT controller consists of a fault detector, a current synchronization unit, and a voltage synchronization unit. As shown in Fig. 5, the proposed FRT controller does not act under normal operating conditions. Under fault conditions, it modifies the current feedback and the power reference to ensure critical FRT functionalities such as maintaining effective synchronization during the fault, supporting reactive power to the faulty phases, preventing reverse active power flow, and maintaining normal operation in the healthy phases.

***B.* <u>The Fault Detector</u>:** The Fault detector, as shown in Fig. 5 is the outermost part of the proposed control architecture. It works as an observer to detect a fault and to activate the proposed FRT controller. From Section III, III.B It is clearly shown that a delay in activating the FRT controller results in a greater deviation of the load angle from the pre-fault value. As a result, the speed of the PLL in the fault detector is crucial for any FRT controller. A PLL model, as presented in [26], is used in the fault detector of the proposed FRT controller. The mentioned PLL model can successfully detect AC voltage amplitude and phase jump of the connected network under the time of half an ac cycle. The PLL model also filters out the dc bias and harmonics.

***C.* <u>Voltage Synchronization when all Three Phases are Under Fault</u>:** The output currents of the grid-forming source cannot be used as feedback for the VOC to establish effective synchronization when all three phases are under fault. The proposed FRT controller uses a PLL-assisted voltage synchronization technique for three-phase faults. The unit vectors of the system voltages, $\hat{v}_{PCCx}$; $(x \in a, b, c)$ are estimated using the PLL. Next, the feedback to the virtual oscillator, $fb_x$; $(x \in a, b, c)$ is modified as follows.

$$fb_x = \gamma(v_{VOCx} - V_{VOCn}\hat{v}_{PCCx});\ (x \in a, b, c) \tag{10}$$

Here, $\gamma$ is a positive scalar which determines the convergence speed [27]. The value of the parameter, $\gamma$ in the proposed FRT controller is selected using the method presented in [27].

***D.* <u>Current Synchronization when at least One Phase is Under Fault</u>:** The current synchronization method is used if at least one phase is healthy during a fault. The current synchronization ensures normal operation in the healthy phase during the fault. In the current synchronization method, the feedback for the faulty phases is estimated using the output currents of the healthy phases as follows.

The three phases, $p_1, p_2, p_3$ have the predefined sequences as

$$p_1 = p_2\angle 120° = p_3\angle 240° \tag{11}$$

During two possible fault conditions, the feedback to the VOC is estimated as follows.

**(i) <u>Condition 1</u>: One of the phases is under fault:** The feedback, $fb_k$ for the $k^{th}$ $(k \in 1, 2, 3)$ phase which is under fault is derived as

$$fb_k = -\sum i_{outx}\,; (x \in 1, 2, 3), (x \neq k) \tag{12}$$

**(ii) <u>Condition 2</u>: Two of the phases are under fault:** The $z^{th}$ phase, $p_z$ is the healthy phase where $z \in 1, 2, 3$. Now, the two faulty phases can be defined as $p_{z+m}$ and $p_{z-m}$ where the limits of $z$ and $m$ is given as

$$m \in 1, 2\ and\ z + m \in 1, 2, 3\ and\ z - m \in 1, 2, 3 \tag{13}$$

Hence, the feedbacks to the faulty phases are derived as

$$fb_{z+m} = i_{outz}\angle -(m \times 120°) \tag{14}$$

$$fb_{z-m} = i_{outz}\angle(m \times 120°) \tag{15}$$

## V. Validation of the Proposed Controller

The simulation from Section II is re-conducted with the proposed unified fault ride-through controller. The waveforms of the main performance parameters during the fault without and with the proposed FRT controller are presented side-by-side in Fig. 6. The same scale is maintained in Fig. 6 for both cases to present a fair comparison.

As expected and as shown in Fig. 6, during the fault $t = t_1$ to $t_2$ , the output current of the grid-forming current is saturated to the maximum current for both the cases, i.e., without and with the proposed FRT controller. However, the main improvement in performance can be seen from Fig. 6. (e) and (f). Fig. 6. (e) shows that the value of $sin\,\delta$ is severely deviated from the pre-fault value without the proposed FRT controller, which indicates a loss of effective synchronization. At the same time, as shown in Fig. 6. (f) value of $sin\,\delta$ is maintained very close to the pre-fault value using the proposed FRT controller.

As shown in Fig. 6. (a) and (b), the mentioned performance improvement helps the grid-forming source to successfully return to a normal operating condition after the fault clearance with a significantly faster speed. Fig. 6. (d) confirms that the proposed FRT controller successfully prevents the reverse flow of active power after the fault clearance.

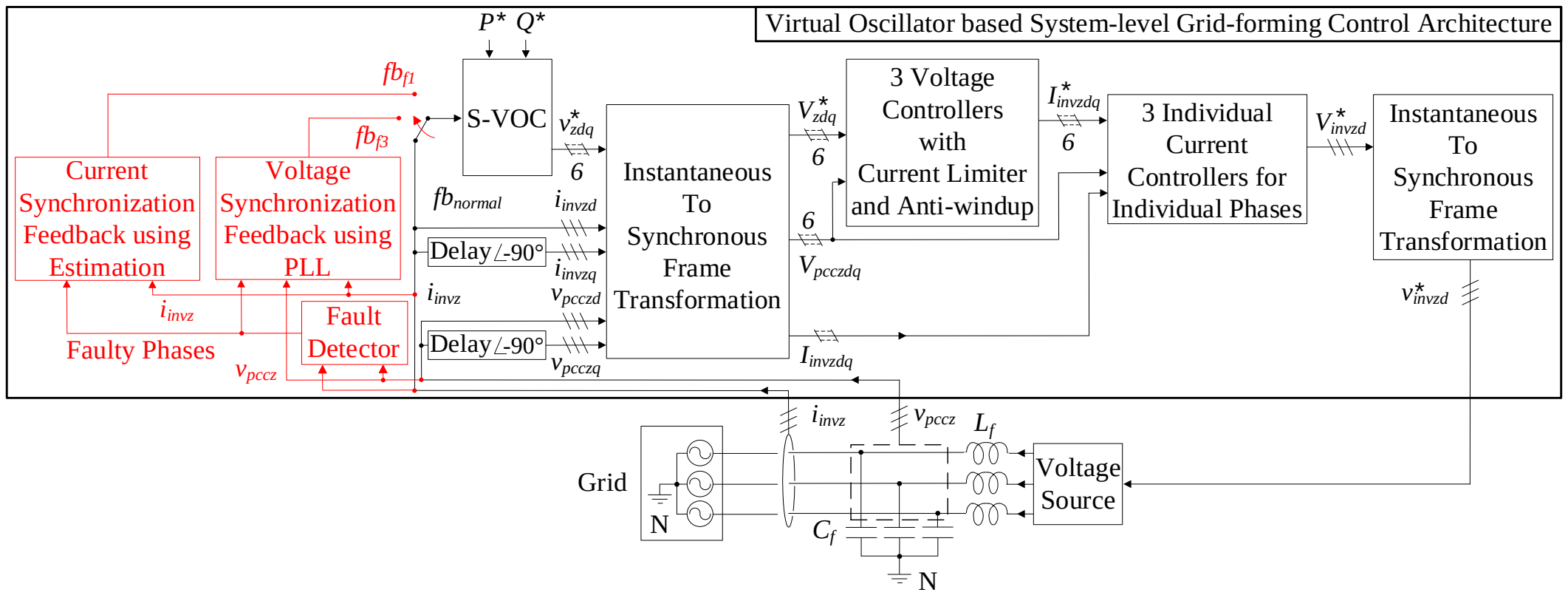

Fig. 5. Integration of the proposed unified fault ride-through controller into a system-level virtual oscillator based grid-forming control architecture

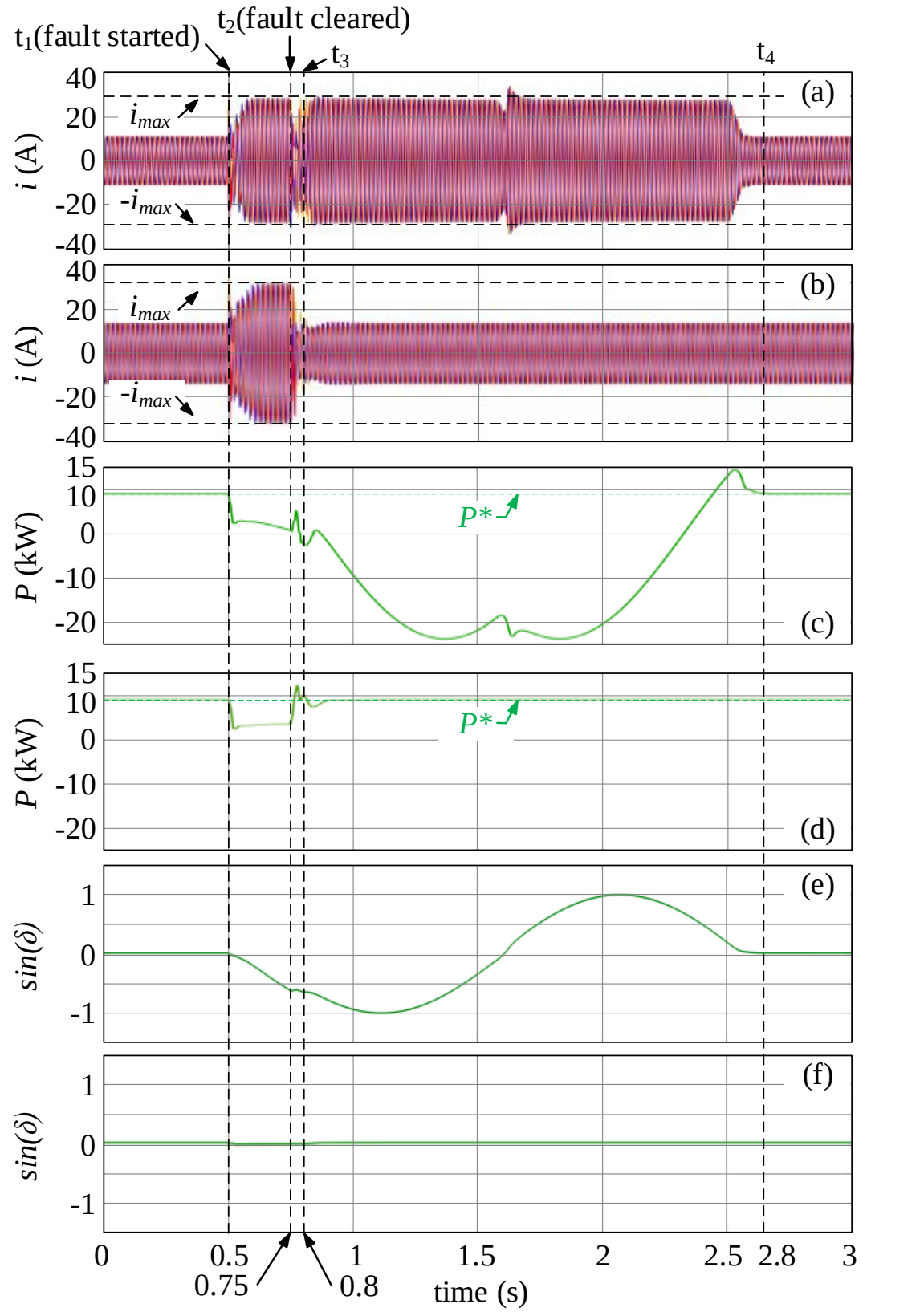

Fig. 6. Fault ride-through performance comparison between existing VOC and existing VOC with the integrated proposed Fault Ride-Through (FRT) controller, (a) and (b): Output current of the grid-forming source without and with the proposed FRT controller; (c) and (d): Active power output of the grid-forming source without and with the proposed FRT controller; (e) and (f): Sine of load angle, $sin\,\delta$ without and with the proposed FRT controller;

Next, the performance of the proposed controller is investigated under unbalanced fault conditions. The active power reference, $P^*$ is set to 15 kw. As shown in Fig. 7 and Fig. 8, prior to the fault, the grid-forming source delivered 5 kw of active power at each phase. During the unbalanced fault conditions, the Current Synchronization Feedback block, as depicted in Fig. 5 is activated. From Fig. 7. (a) and Fig. 8. (a) it is evident that the controller successfully limits the output current of the faulty phases. As presented in Fig. 7. (b) and Fig. 8. (b) the proposed controller prevents any possible active power swing or reversal at the faulty phases. At the same time, the operation at the healthy phases remains normal. After the fault is cleared, the normal steady-state operation is successfully restored at all phases.

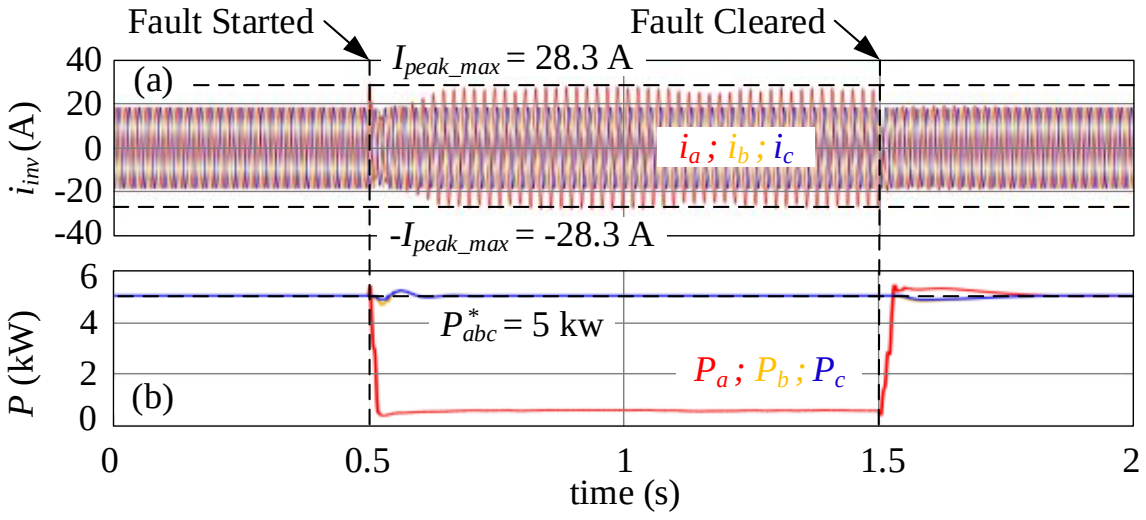

Fig. 7. Fault ride-through performance of the proposed controller under a single line to ground fault, (a): Output current of the grid-forming source; (b): Active power output of the grid-forming source

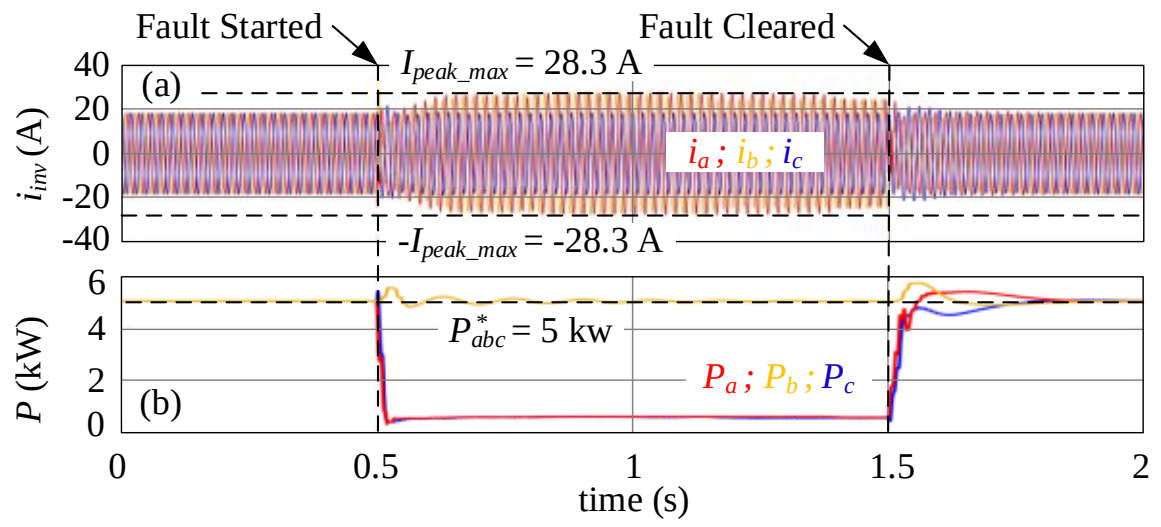

Fig. 8. Fault ride-through performance of the proposed controller under a double line to ground fault, (a): Output current of the grid-forming source; (b): Active power output of the grid-forming source

## VI. Conclusion

This article has focused on addressing critical Fault Ride-Through (FRT) limitations of existing Virtual Oscillator based Controllers (VOCs). This article started with a FRT simulation study and a unique analysis. The mention study identified that the existing VOCs, while effective in limiting output currents during faults, suffer from loss of synchronization with the network, leading to active power swing, active power reversal, and eventual disconnection from the grid. Next, an analytical method is developed to mathematically quantify the interactions between current limiters and the virtual oscillator, contributing to the abovementioned FRT issues. The same mathematical framework provides the essential criteria for the successful FRT operation of VOCs. Finally, a newer FRT control architecture is developed and integrated with existing VOCs without requiring any significant change or additional sensors. The proposed control architecture successfully unified current and voltage synchronization assistance for a grid-forming controller to ride through either balanced or unbalanced fault scenarios.

The performance of the proposed FRT controller is rigorously evaluated, demonstrating its superiority over existing VOCs, meeting critical FRT requirements. Under different fault conditions, the proposed controller has successfully demonstrated the ability to maintain effective synchronization, prevent active power swings and reversal, and ensure normal operation in healthy phases.

The research work, presented in this article, has primarily focused on Virtual Oscillator based Grid-forming controllers. However, the same controller architecture and mathematical analysis can be used to develop advanced FRT control architecture for other types of grid-forming controllers such as Droop Controller and Virtual Synchronous Machine Controller with modifications. In future works, the authors will focus on developing a universal FRT control framework and criteria that is compatible with different types of grid-forming controllers.